\documentclass[12pt,preprint]{aastex}
\newcommand{\be}{\begin{equation}}
\newcommand{\ee}{\end{equation}}
\newcommand{\bbb}{\begin{eqnarray}}
\newcommand{\eee}{\end{eqnarray}}

\newcommand{\filter}{\cal F}

\def\note #1]{{\bf #1]}}
\def\gwig{{\leavevmode\kern0.3em\raise.3ex\hbox{$>$}
\kern-0.8em\lower.7ex \hbox{$\sim$}\kern0.3em}}
\def\lwig{{\leavevmode\kern0.3em\raise.3ex\hbox{$<$}
\kern-0.8em\lower.7ex \hbox{$\sim$}\kern0.3em}}
\slugcomment{Scheduled to appear in the ApJ 20 July 2006, v646n 1 issue}
\begin{document}
\shorttitle{Oscillation amplitudes and travel times}
\title{Sensitivity of time-distance helioseismic measurements to spatial variation of oscillation amplitudes I. 
Observations and a numerical model}
\author{S.P. Rajaguru\altaffilmark{1}, A.C. Birch\altaffilmark{2}, T.L. Duvall Jr.\altaffilmark{3}, M.J. Thompson\altaffilmark{4}, and J. Zhao\altaffilmark{1}}
\altaffiltext{1}{W.W. Hansen Experimental Physics Laboratory, Stanford University, Stanford CA 94305, USA;\email{rajaguru@sun.stanford.edu}}
\altaffiltext{2}{Colorado Research Associates, NorthWest Research Associates, Inc., 3380 Mitchell Lane, Boulder, CO 80301.}
\altaffiltext{3}{Laboratory for Solar and Space Physics, NASA Goddard Space Flight Center, Greenbelt, MD 20771.}
\altaffiltext{3}{Department of Applied Mathematics, Hicks Building, University of Sheffield, Sheffield S3 7RH, UK.}

\begin{abstract}

It is well known that  the observed amplitude of solar oscillations is
lower  in  sunspots  than in  quiet regions of the Sun.   We  show that  this  local
reduction  in  oscillation  amplitudes  combined with the phase-speed
filtering procedure in time-distance helioseismic analyses
could be a source of systematic errors in the range of 5 - 40\% in the measured
travel-time anomalies of acoustic  waves around sunspots.  Removing these travel
time  artifacts is  important for  correctly inferring  the subsurface
structure of  sunspots.  We  suggest an empirical  correction procedure
and illustrate its usage for a small sunspot. This work utilizes data
from MDI/{\it SOHO}.

\end{abstract} 

\section{Introduction}
\label{sec:intro}
The stochastic nature of oscillation excitation due to turbulent convection
is one major source of noise (i.e. realisation noise) and systematics 
in helioseismology \citep{jcd02}.
Another source of systematics is spatial
modulation of the waves by active regions and large-scale convection. 
Accounting for and reducing such noise  and systematics is important in  helioseismology,
global \citep{woodard84,duvallharvey86,schou92,libbrecht92} as well as local
\citep{gizonandbirch04}.  Global  helioseismic power spectral analyses
use  long   uninterrupted  observations  towards this end \citep{libbrecht92,jcd02}.
In contrast, determining localized
non-axisymmetric perturbations  inside the Sun -- the main
goal  of local  helioseismic techniques  -- necessarily  involves using
observations of more limited extent in both space and time.  Such a
task may appear more difficult and susceptible to larger uncertainties
and systematics.  The last  two decades have witnessed the development
and  refinement of  a  new  class of  local  techniques, that  include
helioseismic  holography  \citep{lindseyandbraun97}, far-side  imaging
\citep{lindseyandbraun00}   and  time-distance   helioseismology
\citep{duvalletal93}, which  have been fairly  successful in achieving
such tasks.  These new techniques  are based on  studying quantifiable
properties of causal connections that acoustic waves establish between
points on  the solar surface during  their travel
inside  the  Sun.   These techniques  are  being  fine-tuned,
achieving increased sensitivity to  local changes in the structure and
dynamics      of      the       Sun,      e.g.     sunspots
\citep{sashaetal00,birchetal04,couvidatetal04,hughesetal05,lindseyandbraun05}.
At the same time,
the necessity to  accurately estimate the errors and systematics in the
measurements      is       also      being   increasingly   felt      
\citep{braunetal04,werneetal04,gizonandbirch04}.

Time-distance helioseismology uses temporal cross-correlations of the
oscillation  signals  from  separated  points  on  the  solar  surface
\citep{duvalletal93}.  The wave packet-like structure of such temporal
cross-correlation signals  is understood to be due  to the propagation
of  wave packets   formed  by   acoustic  waves. The waves constituting a
single wave packet travel with
approximately the same horizontal phase speed and in the high frequency
limit follow the same path
inside the  Sun \citep{duvalletal93,duvalletal97}.  Connection between
the time-distance and modal  frequency-wavenumber analyses have
been studied by  \citet{sashaduvall97} and \citet{bogdan97}.
Kosovichev \& Duvall 
provided an  useful formula  to fit  the time-distance
correlation  signals.
Bogdan showed,  with  an  explicit
calculation, that a group of  acoustic waves  with approximately  the same
horizontal  phase-veocity indeed  interfere constructively  to  form a
wave packet  thereby leading  to  the observed  structure of  temporal
cross-correlation  signals.   This   understanding  led  to  further
refinement      of      time-distance      measurement      procedures
\citep{duvalletal97} that include phase-speed filtering:  three-dimensional Fourier spectra
of data  cubes are  filtered to select waves that travel with approximately
the same horizontal  phase speed and inverted  back to  the  time domain  to
perform the  cross-correlations.  Such phase-speed  filtering not only
improves  the signal-to-noise  in travel  time measurements,  but also
makes possible  measuring travel times at very  short travel distances
(shallow  depths   and  hence  of   the  high  degree   modes).  These
improvements are  crucial in measuring travel times  at each location
keeping the original spatial resolution 
of the data, thereby allowing
tomographic   study   of  localized   structures   such  as   sunspots
\citep[e.g.][]{sashaduvall99,sashaetal00,sashaetal01}. 

In this paper we report on the
identification of  a significant source of systematics  in travel-time
measurements that arises  due to  an adverse  coupling  of localized
strong spatial  modulation of oscillation amplitudes  in sunspots with
the phase-speed filtering procedure. Further, we present a numerical model
that describes this source of systematics. 

\section{Oscillation amplitudes and travel times} 
\label{sec:explain1}

\begin{figure}
\centering
\epsscale{0.80}
\plotone{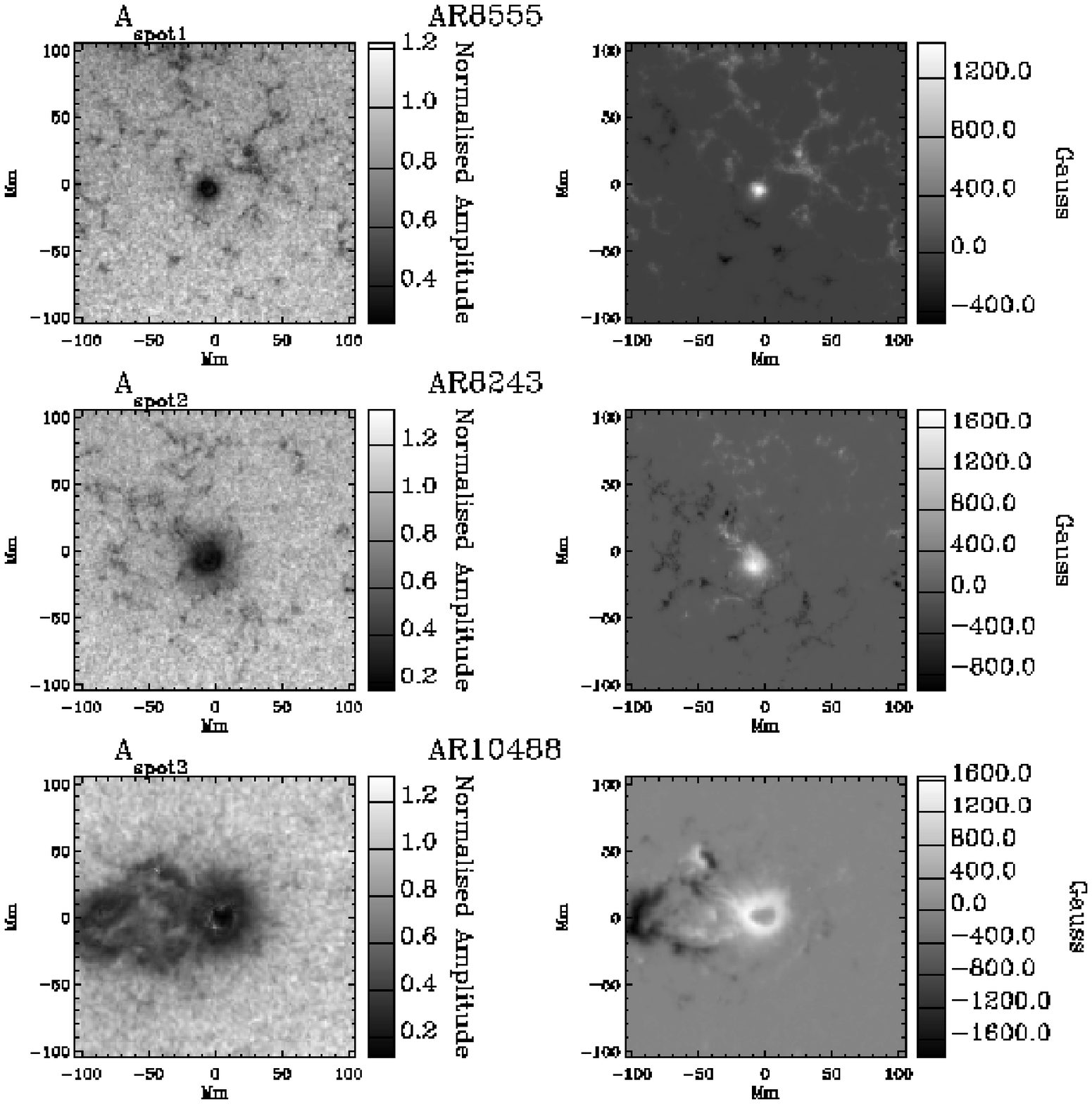}
\caption{Spatial variation of oscillation amplitudes in the p-mode band,
denoted as $A(\mathbf x)$ and defined in the text, normalized to a
quiet-sun spatial average, over small, medium and large
sized sunspot regions as measured from a 512 minute MDI Doppler line of
sight velocity data cubes. The small and medium sized spots' measurements
are from high-resolution, and the large sized spot's is from full disk
resolution MDI data. The right side panels show the corresponding time-averaged
MDI magnetograms.}
\label{fig:1}
\end{figure}

Largely reduced p-mode acoustic  power observed in sunspots is thought
to have contributions  from several causes \citep{hindmanetal97} that
are of two major physical origins: (i) the interaction between sunspot
magnetic field and the quiet-sun  p modes and convection, and (ii) radiative
transfer effects  induced  by altered  thermal  conditions  within the
sunspot. The former physical process is thought to be responsible for (a)
absorption of p modes as known  from a number of studies following the
work of \citet{braunetal88}, (b) alteration of the p-mode
eigenfunctions \citep{jainetal96,hindmanetal97}, and (c) reduced
excitation of p modes within  the sunspot. The latter radiative transfer
effects cause (a) changes in the formation height of spectral lines used 
to measure  the velocities within spots \citep{balthasarandschmidt93,
bogdan00}, and (b) imperfect measurements through changes in the
spectral line profile due to Zeeman splitting and the darkness of the
spot \citep{alamannietal90, wachteretal05}.   The spatial variation of
acoustic  power  can be  determined  from  Doppler  images by  forming
pixel-wise temporal power spectra and  summing the power in the p-mode
band                           of                          frequencies
\citep{hindmanandbrown98,jht_stanchfield00,nicholasetal04}.   Here, we
calculate p-mode power within a  band of frequencies  between 1.7 and
5.3 mHz over three active regions containing a small, medium and large
sized spots using MDI  Doppler velocity data (high-resolution data for
the small and medium size spots, and full-disk resolution data for the
large  spot) \citep{scherreretal95}. The NOAA AR numbers for these three
sunspots are respectively, AR8555, AR8243 and AR10488. We find that quiet-sun regions devoid
of any significant magnetic field show p mode power that
is more or less homogeneous over the solar surface. When a sufficient number of
individual pixel values (or realizations) are averaged over, the p mode power 
is nearly a constant over space in the quiet-sun \citep{venkatetal01}. 
To determine
the relative deviations that active regions introduce in the p mode power 
we normalize the power distribution within  an active
region with respect to a quiet-sun spatial average. The quiet-sun regions
chosen for the normalization
are from within the larger regions covering the sunspots and are of 
the same latitudinal extent as the active regions but are outside of any 
significant magnetic field. We call the square
root  of  such a  spatial  power map  the  oscillation `amplitude  modulation
function', $A(\mathbf x)$, where $\mathbf x$ is the horizontal position
on the solar surface. Note that $A(\mathbf x)$ is derived
by averaging  over the p-mode band  (1.7 -- 5.3 mHz),  but, in general,
amplitude  modulations  are  frequency  dependent. Figure  1  displays
$A(\mathbf x)$ derived for the three active regions chosen, with their
MDI  magnetograms shown as well.  We  note here  that detailed
studies of local magnetic  modulations of oscillation  power and
their  relation  to  the  local  magnetic field  strengths  have  been
reported by  \citet{hindmanandbrown98} and \citet{nicholasetal04}. The
latter authors 
have also constructed  simple models  which allow  a  comparison with
the changes  in  modal power  distribution  determined  from ring  diagram
(normal mode) analyses \citep{rajaguruetal01}.
\placefigure{fig:1}

\begin{figure}[ht]
\epsscale{1.0}
\centering
\figurenum{2}
\plotone{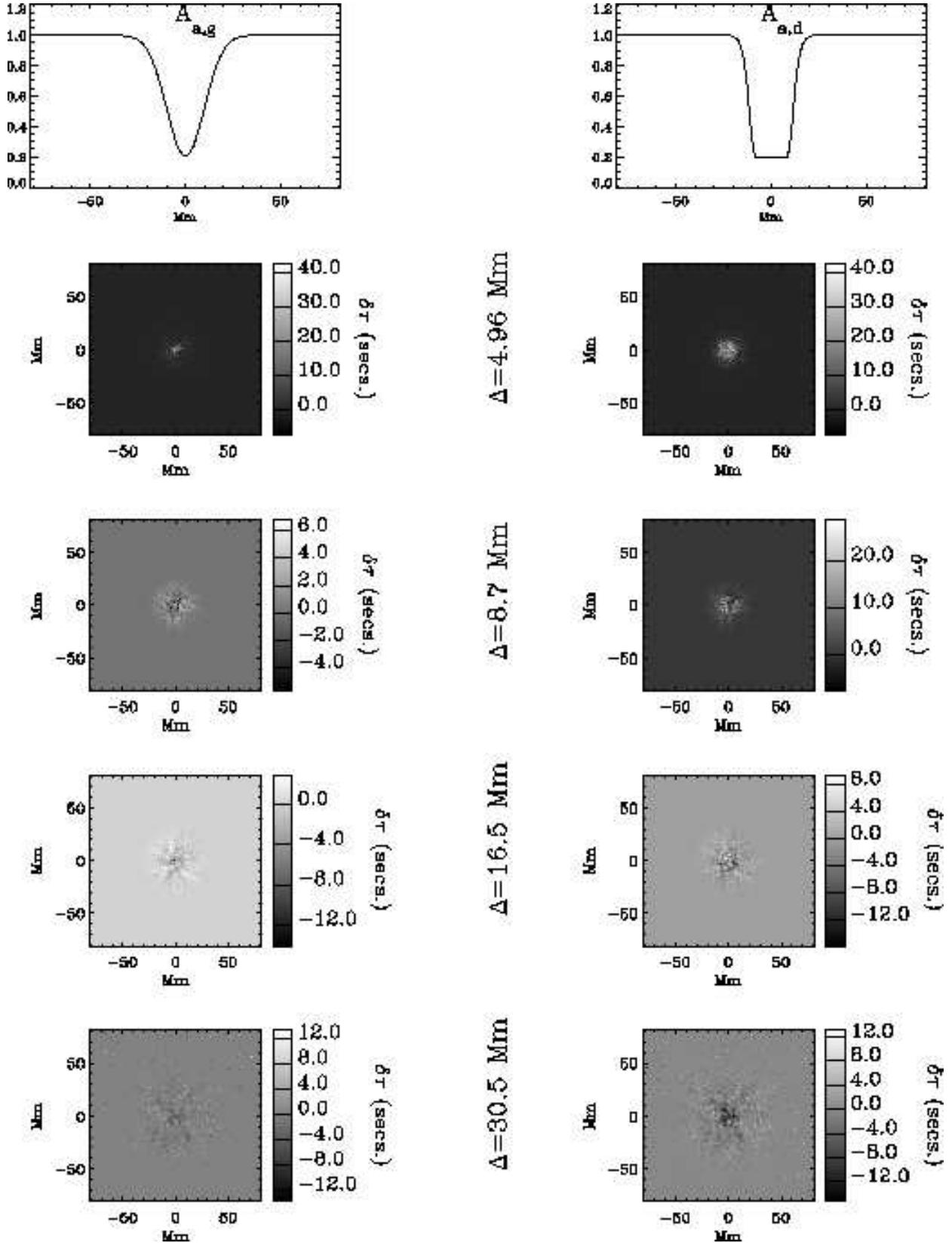}
\caption{ Changes in mean phase travel times, $\delta\tau$, introduced
by oscillation amplitude modulation (spatial) functions $A_{\rm a,g}(\mathbf x)$:
one dimensional cuts through the center of these azimuthally symmetric 2-d
functions are shown in the top row, and $\delta\tau$ are shown below
them for a range of $\Delta$, that are marked between the panels.}
\label{fig:2}
\end{figure}

The measured  spectrum of  oscillations in the  presence of  such long
lived spatial modulation of  oscillation amplitudes is the convolution, in wavenumber space,
of the frequency-wavenumber spectrum of oscillations with the wavenumber
spectrum  of the modulating function. The  possible errors that
such  convolutions would introduce  in the  modal parameters  could be
reduced by using an observational time series sufficiently longer than
the life  span of  amplitude modulators.  Such  a way of  reducing the
systematics is not available in local helioseismology, where
the objective is  to probe perturbations localized in  space and time.
However, a purely {\it time-space analysis} of the oscillation field, in
contrast  to a  {\it frequency-wavenumber analysis}, would  not be
subject to  the kind  of errors from which  a modal power-spectral analysis
suffers.    For  example,  in   time-distance  analysis,   a  temporal
cross-correlation of  oscillation signals  from two locations is not
affected by  a stationary scaling of oscillation  amplitudes and hence
the (phase) travel times  are not affected.  However, the intermediate
step of  phase-speed filtering,  with recourse  to Fourier  space to
select waves of  certain modal relations as explained  in the previous
section, couples the scales or wavenumbers of the modulating function
to the oscillation spectra.  This causes perturbations in the wavenumbers
of the oscillation spectra which in turn manifests as perturbations in
travel times  measured over  regions where the  oscillation amplitudes
are modulated.
\placefigure{fig:2}

\begin{figure}
\epsscale{0.90}
\centering
\figurenum{3}
\plotone{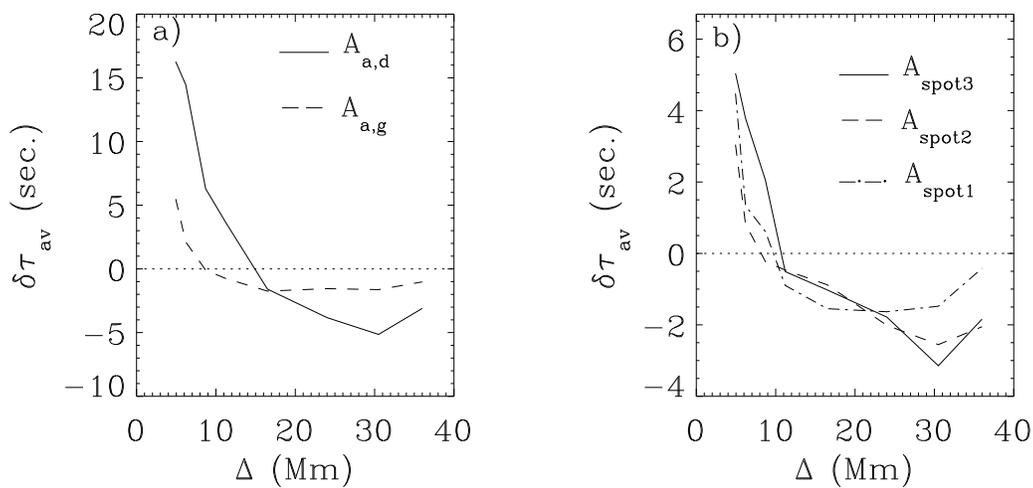}
\caption{The changes in mean phase times spatially averaged over
the masked area, $\delta\tau_{\rm av}$, as a function of travel distance
$\Delta$. Panel $a)$: for amplitude functions $A_{\rm a,g}(\mathbf x)$, and the spatial
area averaged over is 16~Mm in diameter around the peak suppression. Panel $b)$: for
$A(\mathbf x)$ derived from sunspot regions shown in Figure 1 and spatial averages are over the
area of sunspots as seen in the magnetograms.}
\label{fig:3}
\end{figure}

Before  examining this  effect  by way  of  a numerical  model of  the
measurement procedures in the next section, we first demonstrate the changes
in  travel times as  measured using a standard  time-distance analysis
procedure using MDI velocity  data cubes. We perform experiments
using  artificial  amplitude  modulation functions  $A_{\rm a}(\mathbf  x)$,
which bring  out the  essential features of  the coupling  between the
spatial variation  of oscillation amplitudes and the  frequency-wavenumber 
spectrum  of the phase-speed filter.
We choose two forms for $A_{\rm a}(\mathbf  x)$ for this purpose; horizontal 
one-dimensional cuts across these
modulation functions  are shown in the  top row of Figure  2a. We have
chosen a  peak suppression  of 80\% (which is  typical of  umbrae of
medium sized spots, see Figure 1) for both functions. The Gaussian
form, denoted as  $A_{\rm a,g}$ (top left panel in Figure  2a), has a FWHM
of about 16 Mm while the disc-like function, denoted as $A_{\rm a,d}$
(top  right  panel  in  Figure  2a),  has  a  sharp  spatial  gradient
connecting zero  suppression to the peak suppression,  which is spread
over a  disc of diameter  16 Mm. Each  velocity image of a  very quiet
region data cube is multiplied  by $A_{\rm a}(\mathbf x)$ before running the
data through a standard time-distance analysis procedure that includes
phase-speed filtering  and uses center-annulus  geometry for computing
cross-correlations  \citep{duvalletal97,rajaguruetal04}.  
Travel time maps are  calculated for the range in travel
distances $\Delta$  that are normally used  in tomographic inversions
\citep{sashaduvall99,couvidatetal04,hughesetal05},  and  are  compared
with  those  obtained  for   the  original  quiet-sun  data  (without
introducing  any  amplitude variations).   The  shifts  in mean  phase
travel  times, i.e.  mean of  ingoing and  outgoing wave  phase travel
times,  $\delta\tau_{\rm  mean}=\tau_{\rm mean}(\rm  masked)-\tau_{\rm
mean}(\rm  quiet)$, as a  function of  $\Delta$ are  shown as  maps in
Figure 2. Hereafter, by  travel times we always refer to mean
phase travel times  and remove the subscript 'mean'  in the notations,
i.e.  $\delta\tau  =  \delta\tau_{\rm  mean}$. Figure  3  shows  these
$\delta\tau$ spatially  averaged over the masked area,  which is about
16 Mm in diameter, and denoted as $\delta\tau_{\rm av}$, as a function
of  $\Delta$. The  results in  Figure  2 and  3a show  the
following main features  of amplitude suppression on travel times. Firstly,
steeper  spatial gradients  in  the  amplitude suppression  cause
larger  shifts in  travel times  (compare  left and  right columns  in
Figure 2) in addition  to the proportional  changes caused by
the  amount of suppression. Secondly,  smaller $\Delta$  show positive
shifts in travel times (longer travel times), while the larger $\Delta$ show the
opposite change (shorter travel times), with the change over occurring at larger $\Delta$ for
larger   spatial   gradient  suppression. Thirdly, the  magnitude   of
$\delta\tau_{\rm av}$ decreases as $\Delta$ increases.
\placefigure{fig:3}

\begin{figure}[ht]
\epsscale{0.90}
\centering
\figurenum{4}
\plotone{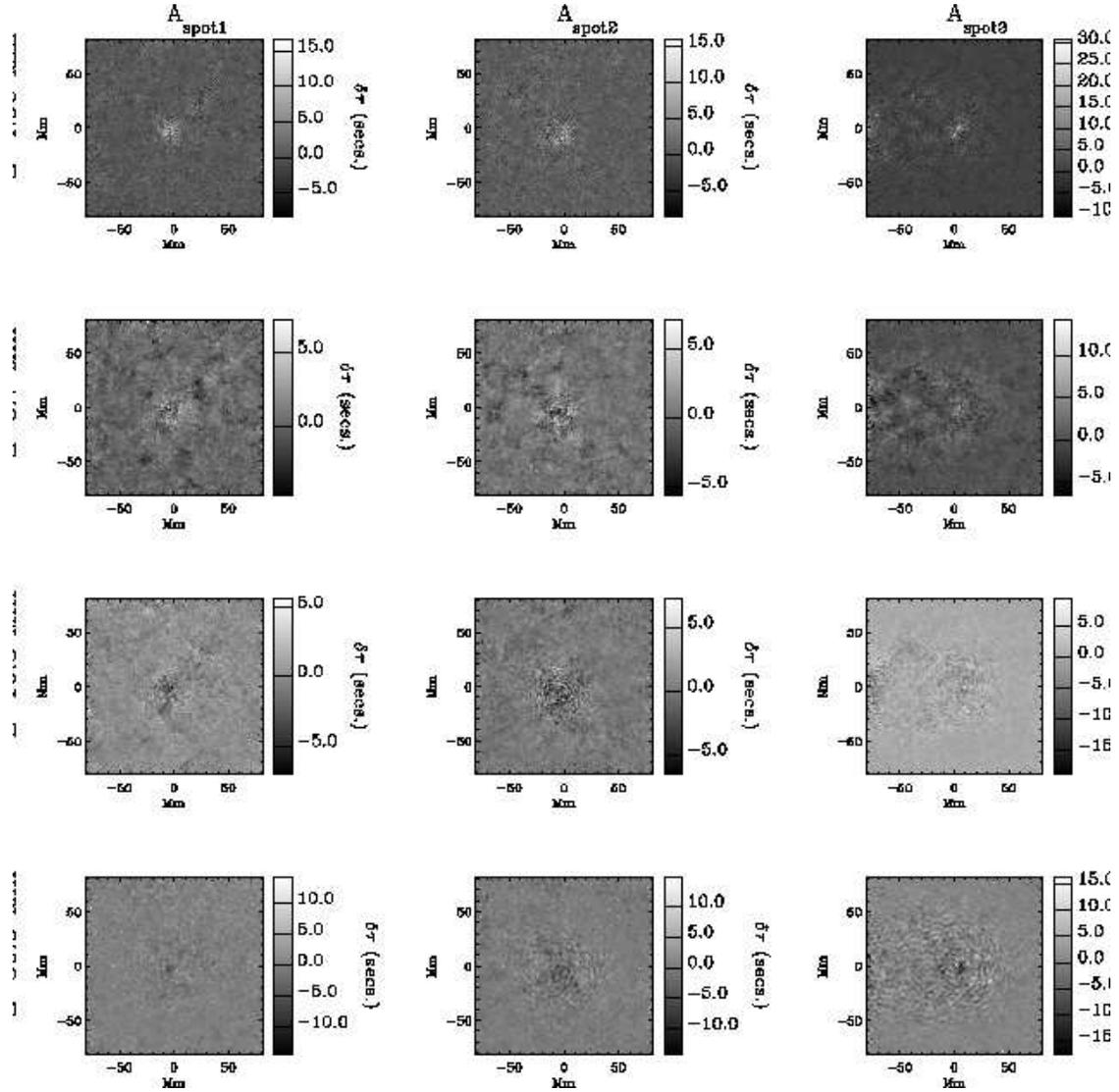}
\caption{Same as in Figure 2 but for amplitude functions
$A(\mathbf x)$ derived from sunspot regions that are shown in Figure 1.}
\label{fig:4}
\end{figure}

To  estimate changes  in travel  times that  sunspots  could introduce
purely due to the spatial variation that they cause in the oscillation
amplitudes,  we   then  apply  $A(\mathbf  x)$   determined  from  the
pixel-wise power map as explained earlier and shown in Figure 1 to the
same  quiet-sun  patch  data cube  and  compare the  travel-time  maps
obtained  with and  without  the application  of  $A(\mathbf x)$.  The
results for $\delta\tau$  are shown in Figure 4, similarly to
that shown  in Figure 2. Figure 3b compares the
$\Delta$  dependence  of  $\delta\tau_{\rm  av}$, which  are  averaged
$\delta\tau$ over the surface area  of the spots, for the 
small, medium and large sized spots.
\placefigure{fig:4}

\begin{figure}
\figurenum{5}
\centering
\plotone{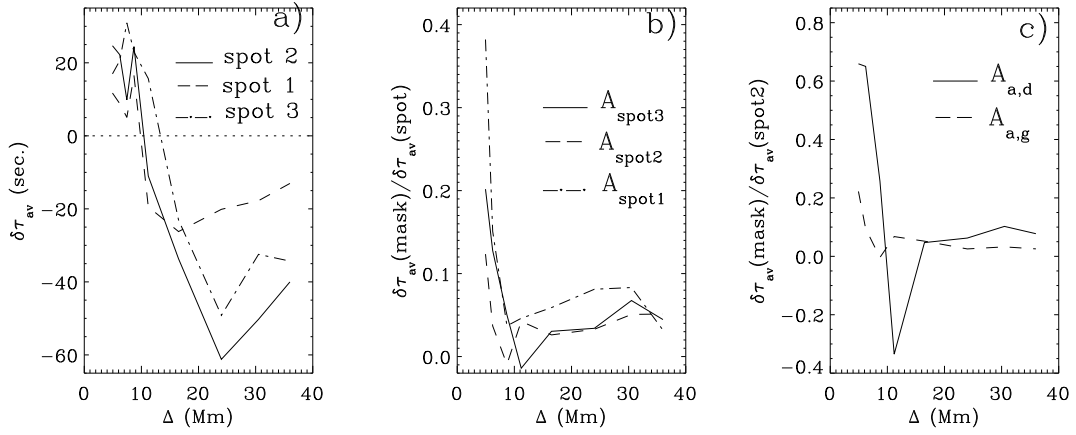}
\caption{Panel a): spatial average of $\delta\tau=\tau({\rm spot})-\tau({\rm quiet})$ measured
over the three sunspots (Figure 1) as a function of $\Delta$. Panel b): the
fractional values of the average shifts, shown in Figure 3b, with respect to those
shown in panel a) here for the three spots; and in panel c) are those in
Figure 3a scaled by that of medium sized spot (spot 2).}
\label{fig:5}
\end{figure}

How do the changes we have measured  and shown in Figures 2 -- 4, which
are purely due  to the combined action of  spatial amplitude variation
and  phase-speed filtering,  compare  with  the  actual travel  times
measured in sunspot  regions? For this purpose, we  have measured the
mean travel-time shifts over the three sunspot regions shown in 
Figure  1 with exactly  the same measurement procedure
as  for the results shown  in  Figures 2  --  4.   The  $\Delta$
dependence of  $\delta\tau=\tau (\rm spot)-\tau  (\rm quiet)$ averaged
over  the area of  the spots  is shown  in Figure  5a.  The
fractional values,
 with respect  to the $\delta\tau_{\rm av}$ measured
for  these spots,  of the  similar changes  measured over the amplitude
modulated areas shown in Figure 3 are shown in panels $b)$ and $c)$ of
Figure 5. In summary, the results in Figures 2 -- 5 show that spatially
localized  amplitude variations  in  the oscillation  field caused  by
sunspots (Fig. 1), in combination  with the phase-speed filtering in
the  analysis procedure,  can account for mean travel-time shifts in the
range of 5 -- 40\% in  the
observed travel-time anomalies in sunspots.  
\placefigure{fig:5}

A simple experiment of applying the amplitude modulation after the
phase-speed filtering leads to negligible changes in travel time. This
proved to us that the effect was caused by the interaction of the 
phase-speed filter with the amplitude modulation. A  clear understanding of
the origin,  and a  method of  accounting for it  in the  travel times
measured in sunspots, of such  changes are important because these can
be a source of systematic  errors in the subsurface inferences derived
using differential inversion methods such as are described in
\citet{gizonbirch05} and references therein. In the next section we build a 
numerical model of the action of phase-speed filter and its interaction 
with an amplitude function $A(\mathbf  x)$. 

\section{Action of a Phase-Speed filter} 
\label{sec:explain2}
In this  section we derive  a simple model  showing the effect  of the
amplitude suppression function $A(\mathbf x)$ on the center-to-annulus
cross-covariance  and hence  the center-to-annulus  travel  time.  The
three generic  steps in computing  center-to-annulus cross-covariances
are to filter  the data, average the data over the  annulus and over a
small region  around the  center point to  obtain the  ``annulus'' and
``center'' signals,  and then to  compute the cross-covariance  of the
``center'' and ``annulus''  signals \citep{duvalletal97}. We write the
observed oscillation signal, $\phi$,  e.g.\ the line-of-sight component
of the velocity at the solar surface as,
\begin{equation}
\label{eq:amplitude}
\phi(\mathbf x,t) = A(\mathbf x) \psi(\mathbf x,t) \, ,
\end{equation}
where $\psi(\mathbf x,t)$ is the underlying oscillation signal, i.e. the signal that would be
seen if the sunspot had no effect on the oscillation amplitude.  
Horizontal position is given by ${\mathbf x}$ and time by $t$.

The first step is to filter the observed signal.  
This is
performed in the Fourier domain (${\mathbf k, \omega}$) by multiplying the
Fourier transform of the observed signal, $\phi(\mathbf k, \omega)$, by
a filter function $\filter(\mathbf k, \omega)$:
\begin{equation}
\phi_{\rm f}(\mathbf k, \omega) = \filter(\mathbf k, \omega) \phi(\mathbf k, \omega)
\end{equation}
This filtering can be transformed back from wavenumber to the
space domain as a convolution over space:
\begin{equation}
\label{eq:filter_action0}
\phi_{\rm f}(\mathbf x, \omega) = \filter(\mathbf x, \omega) \otimes \phi(\mathbf x, \omega)
\end{equation} 
where $\filter(\mathbf x, \omega)$ is the inverse Fourier transform over
wavenumber $\mathbf k$ of the filter function $\filter(\mathbf k, \omega)$ and
$\otimes$ is a convolution over space $\mathbf x$.  Using the convention
for discrete Fourier transforms given in Appendix~1 of Gizon \& Birch (2004),
Eq.~\ref{eq:filter_action0} can be rewritten as,
\begin{equation}
\label{eq:filter_action}
\phi_{\rm f}(\mathbf x,\omega) = \frac{h_x^2}{(2\pi)^2}\sum_{\mathbf x'} 
\filter\left(\mathbf x-\mathbf x',\omega\right)\phi(\mathbf x',\omega) \, ,
\end{equation}
where the sum over surface position $\mathbf x'$ is taken over 
all points where $\filter\left(\mathbf x-\mathbf x',\omega\right)$ is 
not zero, and $h_x$ is the grid spacing in the horizontal 
directions (near disk center $h_x=1.39$~Mm for full-disk MDI data 
and $h_x=0.83$~Mm for two-by-two binned high-resolution MDI data).

\begin{figure}
\centering
\figurenum{6}
\plotone{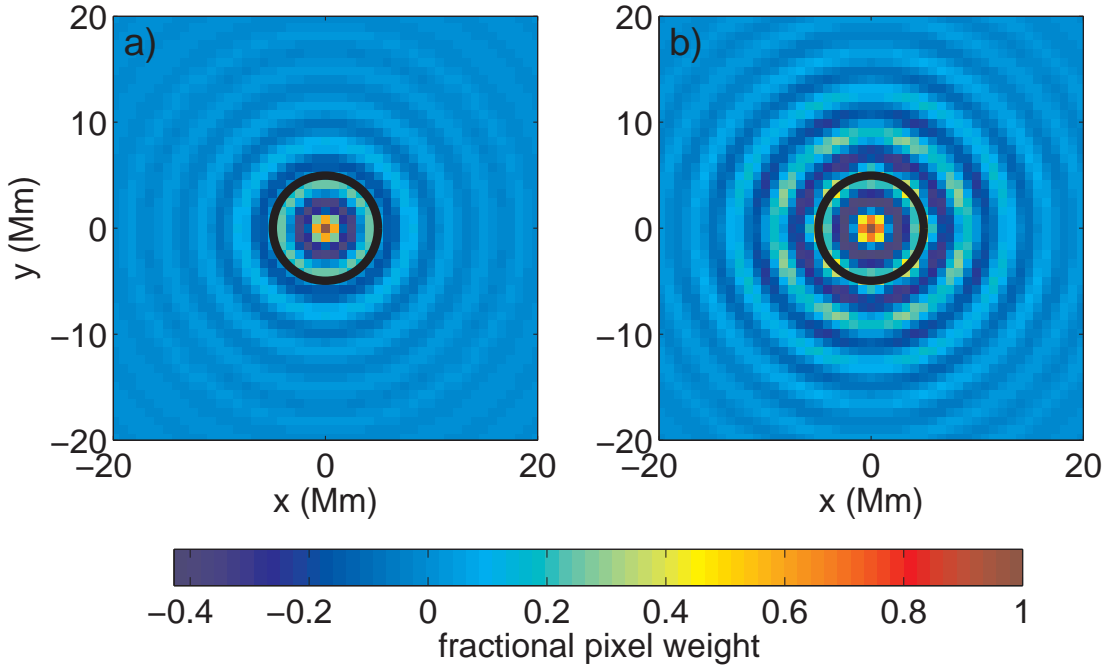}
\caption{Spatial dependence of weight functions at fixed
$\omega/2\pi=3.9$~mHz and distance $\Delta=4.96$~Mm.  Panel a shows $W_{\rm center}$ and panel b shows
$W_{\rm annulus}$.  In both cases the heavy black line shows the nominal
distance, $\Delta$.  }
\label{fig:6}
\end{figure}

The second step is to average the filtered signal over the annulus and 
then, separately, over a small region around the center point to obtain 
the ``annulus'' and ``center'' signals respectively (see \citet{rajaguruetal04} for 
a detailed description of the exact procedure employed in this paper)
In general we can write
\begin{eqnarray}
\phi_{\rm annulus}(\mathbf x,\Delta,\omega) &=& \sum_{\mathbf x'} f_{\rm 
annulus}(\mathbf x'-\mathbf x,\Delta) \phi_{\rm f}(\mathbf x',\omega)  \label{eq:phi_ann} \, 
,\\
\phi_{\rm center}(\mathbf x,\omega) &=& \sum_{\mathbf x'} f_{\rm center}(\mathbf x'-\mathbf x) 
\phi_{\rm f}(\mathbf x',\omega)  \label{eq:phi_center}  \, ,
\end{eqnarray}
where $f_{\rm annulus}(\mathbf x,\Delta)$ and $f_{\rm center}(\mathbf x)$ are the 
weight functions for obtaining the ``annulus'' and ``center'' signals from 
the filtered data \citep{rajaguruetal04}.
Combining the above equations 
(Eqs.~[\ref{eq:phi_ann}-\ref{eq:phi_center}]) with 
equation~(\ref{eq:filter_action}) we obtain
\begin{eqnarray}
\phi_{\rm annulus}(\mathbf x,\Delta,\omega) &=& \sum_{\mathbf x'} W_{\rm 
annulus}(\mathbf x'-\mathbf x,\Delta) \phi(\mathbf x',\omega)  \label{eq:W_ann} \, ,\\
\phi_{\rm center}(\mathbf x,\omega) &=& \sum_{\mathbf x'} W_{\rm center}(\mathbf x'-\mathbf x) 
\phi(\mathbf x',\omega)  \label{eq:W_center} \, ,
\end{eqnarray}
with the weight functions $W$ given by
\begin{eqnarray}
W_{\rm annulus}(\mathbf x,\Delta,\omega) &=& \frac{h_x^2}{(2\pi)^2} \sum_{\mathbf x'} 
f_{\rm annulus}(\mathbf x'+\mathbf x,\Delta) \filter(\mathbf x',\omega) \,  \\
W_{\rm center}(\mathbf x,\Delta,\omega)  &=& \frac{h_x^2}{(2\pi)^2} \sum_{\mathbf x'} 
f_{\rm center}(\mathbf x'+\mathbf x) \filter(\mathbf x',\omega) \, . 
\end{eqnarray}
We have now expressed (Eqs.~[\ref{eq:W_ann}]-[\ref{eq:W_center}]) the 
``center'' and ``annulus'' signals as averages of the unfiltered data.
The weight functions $W_{\rm annulus}$ and $W_{\rm center}$ express the 
weights with which the raw data are averaged, at each
temporal frequency, to obtain these average signals. 
Figure~\ref{fig:6} shows an example of these weight functions.  In 
general,
the weight functions are not well localized as a result of the strong 
horizontal wavenumber dependence of the phase-speed filtering.
\placefigure{fig:6}

The final step is to compute the cross-covariance of the ``center'' and 
``annulus'' signals. This cross-covariance, for center position $\mathbf x$ and 
annulus with radius $\Delta$, is given by
\begin{equation}
\label{eq:cov_filtered_phi_orig}
C(\mathbf x,\Delta,\omega) = 2\pi \phi_{\rm center}^*(\mathbf x,\omega) \phi_{\rm 
annulus}(\mathbf x,\Delta,\omega) \, .
\end{equation}
Employing equations~(\ref{eq:W_ann}) and~(\ref{eq:W_center}) we can write 
equation~(\ref{eq:cov_filtered_phi_orig}) as
\begin{equation}
\label{eq:cov_filtered_phi}
C(\mathbf x,\Delta,\omega) = \sum_{\mathbf x',\mathbf x''} W^*_{\rm center}(\mathbf x'-\mathbf x) 
W_{\rm annulus}(\mathbf x''-\mathbf x) C^\phi(\mathbf x',\mathbf x'',\omega)
\end{equation}
where the point-to-point cross-covariance of $\phi$ is
\begin{equation}
  C^\phi(\mathbf x',\mathbf x'',\omega) = 2\pi \phi^*(\mathbf x',\omega)\phi(\mathbf x'',\omega)\, 
.
\end{equation}
We can express equation~(\ref{eq:cov_filtered_phi}) in terms of the 
covariance of the underlying wavefield $\psi$ by noticing that the 
point-to-point cross-covariance of $\phi$ can be
written in terms of the point-to-point cross-covariance, $C^\psi$, of the 
underlying wavefield, $\psi$, and the amplitude suppression function $A$ 
as
\begin{equation}
  C^\phi(\mathbf x',\mathbf x'',\omega) = 2\pi A(\mathbf x') A(\mathbf x'') 
\psi^*(\mathbf x',\omega)\psi(\mathbf x'',\omega) \, .
\end{equation}
As a result, the center-to-annulus cross-covariance of $\phi$ is
\begin{equation}
\label{eq:cov_filtered_phi_A}
C(\mathbf x,\Delta,\omega) = \sum_{\mathbf x',\mathbf x''} W^*_{\rm center}(\mathbf x'-\mathbf x) 
W_{\rm annulus}(\mathbf x''-\mathbf x) A(\mathbf x')A(\mathbf x'') C^\psi(\mathbf x',\mathbf x'',\omega) \, .
\end{equation}
This is the desired result; we have the center-to-annulus cross-covariance 
of the filtered wavefield in terms of the point-to-point cross-covariance
of the underlying wavefield $\psi$. As described in Gizon \& Birch (2004) 
we  can compute $C^\psi$ in terms of only the power spectrum
of $\psi$.  The weight functions $W$ depend only on the filter function 
and the averaging done to obtain the ``center'' and ``annulus''
signals.  Thus, for a given measurement scheme (i.e. filter and spatial 
averaging scheme) and for a particular  power spectrum of the underlying 
wavefield,
we can compute how the center-to-annulus cross-correlation of the modified 
wavefield depends on the amplitude function $A$.  From 
equation~(\ref{eq:cov_filtered_phi_A})
we can see that the effect of an amplitude function is to alter the 
weights with which different two-point cross-covariances contribute to the
full center-to-annulus cross-covariance.

\begin{figure}
\figurenum{7}
\plotone{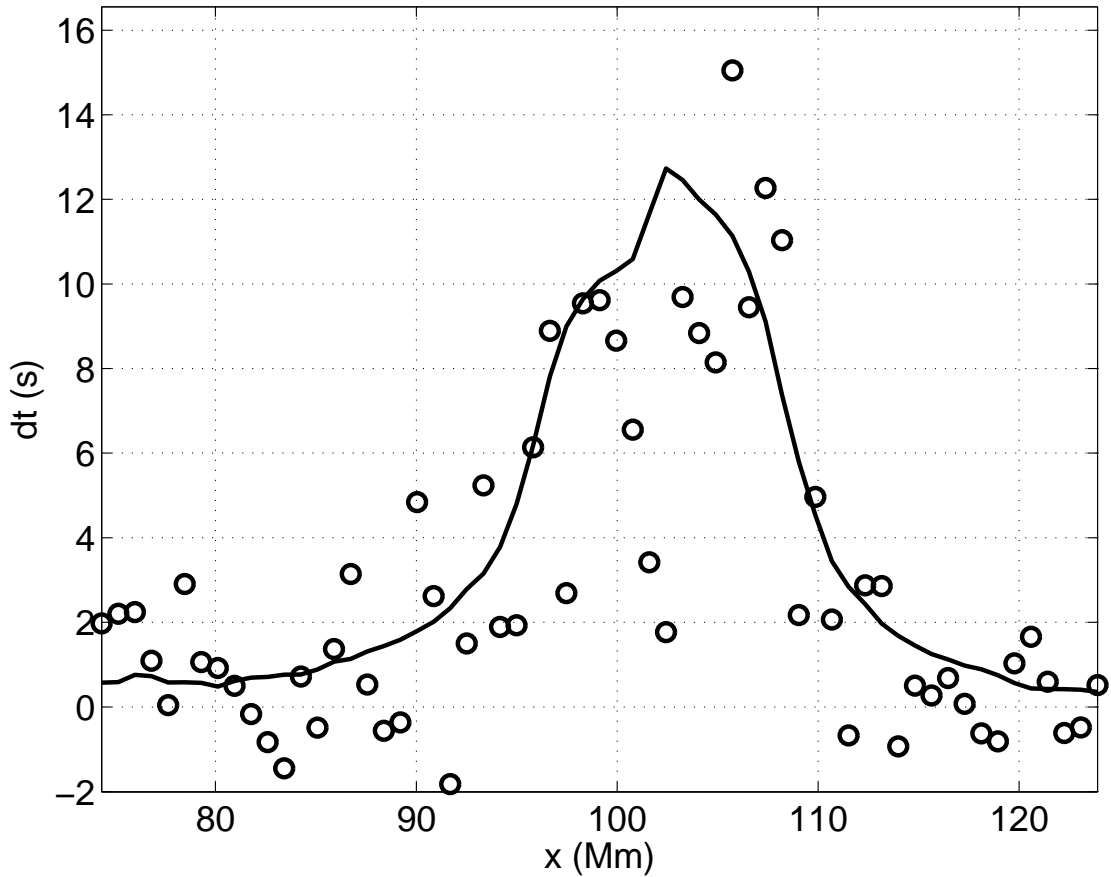}
\caption{Travel time measured from the theoretical cross-covariances
(Eq.~[\ref{eq:cov_filtered_phi_A}]), solid line, and travel times measured
from the masking experiment shown in Figure~4a (top left panel) at $y \approx 98$~Mm. There is good qualitative
agreement within the noise level of the data. }
\label{fig:7}
\end{figure}

\placefigure{fig:7}
In      order      to       demonstrate      the      validity      of
equation~(\ref{eq:cov_filtered_phi_A})       we      computed      the
cross-covariances $C(\mathbf x,\Delta,\omega)$  for the  case shown in the  top left  panel of
Figure~4: this  case corresponds to applying  the amplitude variation
measured over  the small  sunspot (spot  1) to a  quiet-sun  patch and
measuring  the travel-time  shifts for  a distance  $\Delta$ of  4.96 Mm.

In
Figure~\ref{fig:7}  we  show  a  comparison  of  the
travel times shown in Figure~4 (top left panel) with the travel times
measured          from          $C(\mathbf x,\Delta,\omega)$         predicted          by
equation~(\ref{eq:cov_filtered_phi_A}).  We see that the model presented in this
section predicts the effect of the amplitude function $A$ reasonably well.

\section{An empirical correction procedure} 
\label{sec:analysis}

\begin{figure}
\figurenum{8}
\epsscale{1.10}
\plottwo{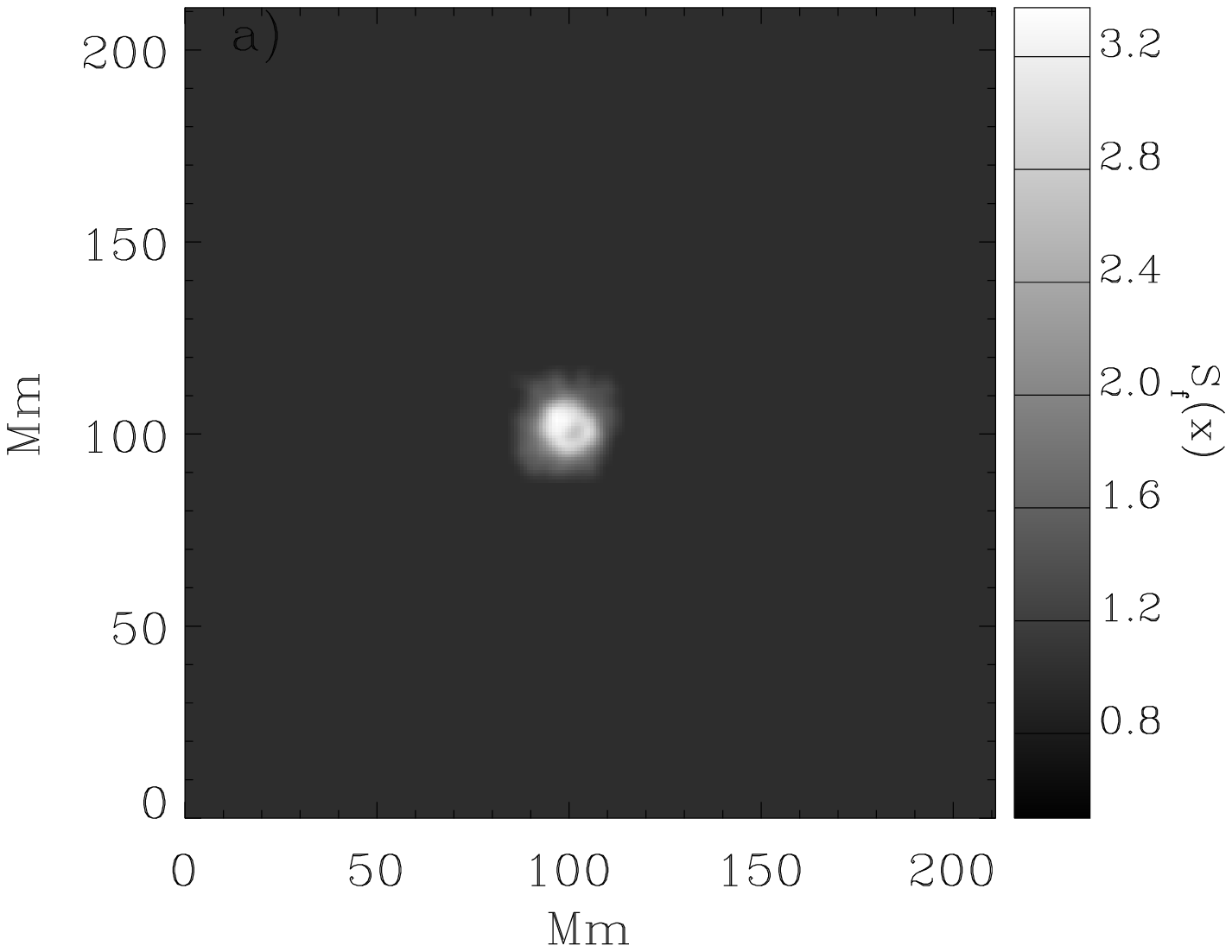}{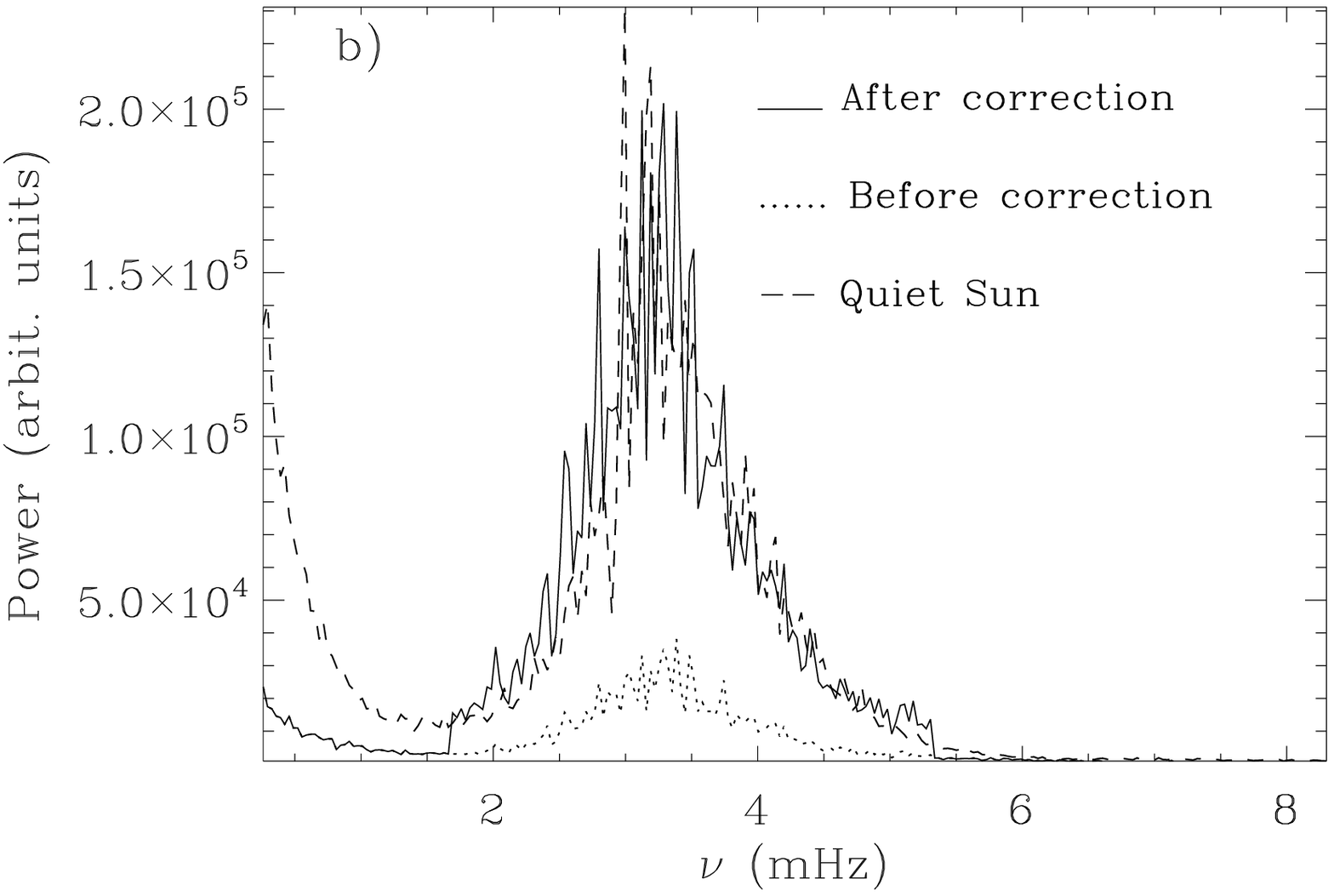}
\caption{Panel $a)$: spatial map of scale-up factor $S_{\rm f}(\mathbf x)$ determined
around the sunspot of Figure 1 (top row). This is the factor by which p-mode amplitudes
are boosted up. The frequency distribution of power averaged over sunspot pixels before
({\it dotted line}) and after ({\it solid line}) the corrections is shown in panel $b)$;
power averaged over a same number of quiet-sun pixels is shown as {\it dashed line}}.
\label{fig:8}
\end{figure}

The results  in the previous  two sections show that systematic
shifts in  travel times are caused  by the interaction  of the spatial
variation  in  oscillation  amplitudes,  be  they of  any  origin  (as
demonstrated by our experiments with artificial modulation functions),
with  the  phase-speed  filtering  in the  analysis  procedure.  This
understanding shows  that if  we could  remove the
strong spatial variation in  the oscillation amplitudes, caused by the
objects  of our  main concern,  i.e. sunspots,  without  affecting the
temporal phase evolution of  oscillation signals, then this particular
effect could be reversed  thereby removing the systematic shifts. This
suggests that the  oscillation signal at each pixel could
be boosted  up by a constant  factor obtained from the
amplitude  function $A(\mathbf  x)$  (Fig.  1)  derived  from  the
pixel-wise power maps: a simple way of estimating the pixel-wise scale-up 
factor is just taking the inverse of $A(\mathbf x)$.
\placefigure{fig:8}

A natural  way to carry  out this remedy  is to boost  the amplitude of the oscillation
signal in each pixel  over the  sunspot region  so that  the functions
$A(\mathbf x)$  look smoother and have  values similar to  that of the
more  or less homogeneous  quiet-sun  regions. We  choose the  case of
the small  sunspot  (spot  1) shown  in  the  top  row  of Figure  1.  The
pixel-wise  scale-up  factor  is  given by  $S_{\rm  f}(\mathbf x)$  =
1/$A(\mathbf  x)$. Here,  we  concern ourselves  with correcting  the
power deficit only  within the sunspot, and so  we determine the scale-up  
factor $S_{\rm  f}$  in a  small  area in  and  around the
sunspot.  The sunspot (spot  1) is found to be about 14 Mm
in diameter, as  seen in $A(\mathbf x)$ (top left  panel in Figure 1),
and we choose an area of  about 28 Mm square centered around this spot
and calculate $S_{\rm  f}(\mathbf x)$ in this region.  To minimize the
effects of pixel-scale variations  we smooth $S_{\rm f}$  by a four
pixel box  car.  The resulting map  of the scale-up factor  is shown in
Figure 8a. We note  that $A(\mathbf x)$ was  determined for p
modes within a frequency band of 1.7  -- 5.3 mHz. Hence we use $S_{\rm
f}$ calculated  as above  to boost  the amplitudes of  p modes  in the
same band  of frequencies,  i.e. we  apply $S_{\rm  f}$  in
Fourier space to boost the amplitudes of p modes in this band. These are then
inverted back  to the  time-space domain  to get  the corrected
Doppler velocity data cube that is subjected to the same time-distance
analysis procedure as before (Section 2). Figure 8b shows the
frequency  distribution of  power  averaged over  the sunspot  pixels
(an area of 14 Mm square around the spot center) before and after correction, i.e. before and after
applying $S_{\rm f}$, and also quiet-sun power averaged over an area of the same size.
We  note   here  that  such  artificial  enhancement  of
oscillation  amplitudes will  not undo  the real  physical  changes in
travel times that the spots have  caused, but are expected to undo the
changes arising from the  amplitude modulations demonstrated
in the previous section.

\begin{figure}
\figurenum{9}
\epsscale{0.90}
\plotone{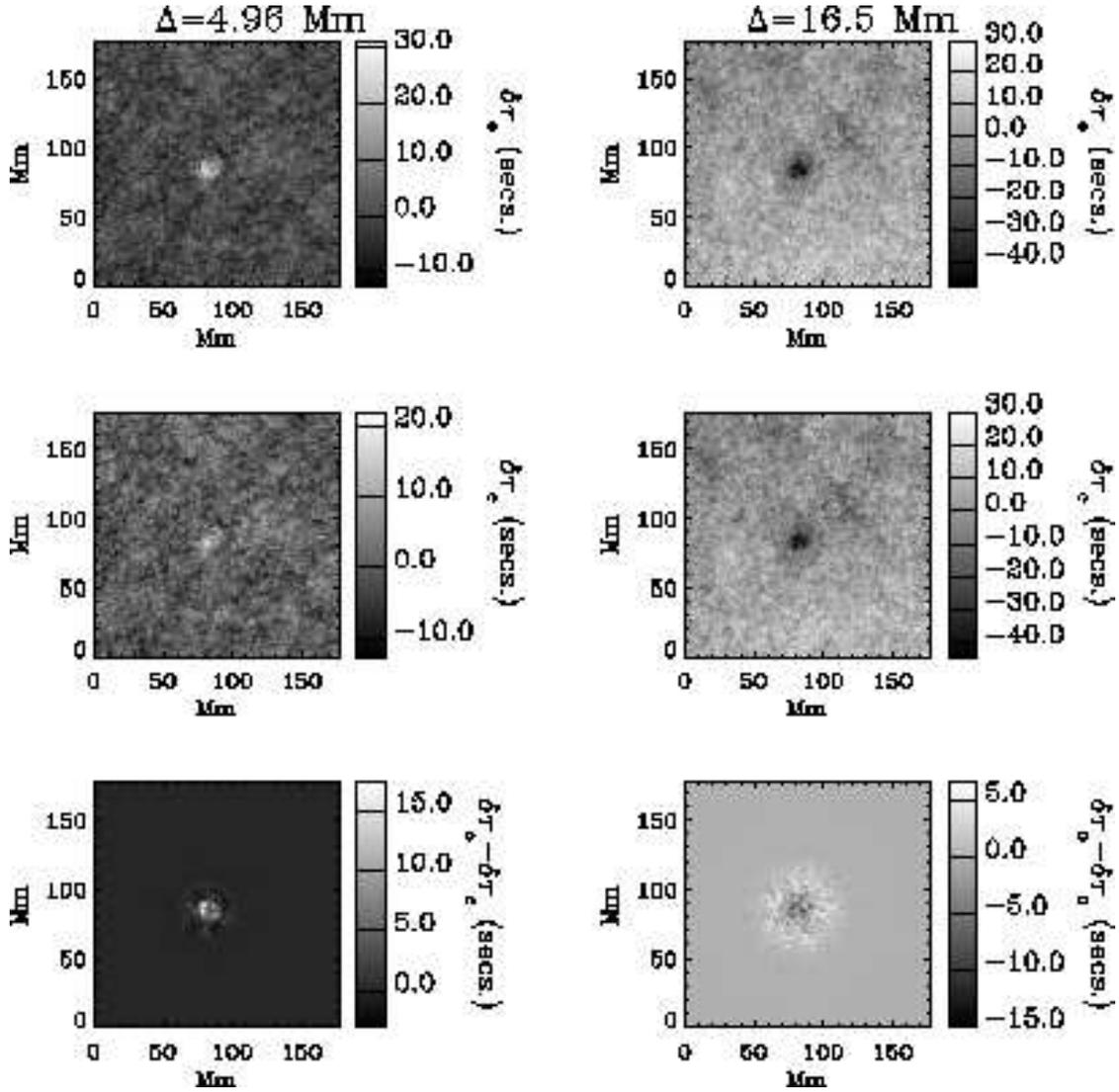}
\caption{Maps of changes in mean phase travel times $\delta\tau$ over the small
sunspot (spot 1) region for two $\Delta$ (in the two columns). In the top row
are the original (uncorrected) travel times $\delta\tau_{\rm o}$, in the middle
row are the corrected ones $\delta\tau_{\rm c}$, and the bottom row shows the
corrections $\delta\tau_{\rm o} -\delta\tau_{\rm c}$.}
\label{fig:9}
\end{figure}

\placefigure{fig:9}
We calculated maps of changes in mean phase travel times, $\delta\tau =
\delta\tau({\rm  spot}) - \delta\tau({\rm  quiet})$, before  and after
the amplitude or power corrections described above. The results for two
representative travel distances $\Delta$ of 4.96 and 16.5 Mm are shown
in  the  two columns  of  Figure  9. The top  row  shows  the original  or
uncorrected travel  times $\delta\tau_{\rm  o}$, the middle  row shows
the corrected travel times $\delta\tau_{\rm c}$, and in the bottom row
are shown the  differences $\delta\tau_{\rm o} - \delta\tau_{\rm  c}$. 
It is instructive to compare $\delta\tau_{\rm  o} - \delta\tau_{\rm c}$ with
the corresponding  panels (i.e.\  for  the same $\Delta$) in  Figures 2
and 4: the corrections $\delta\tau_{\rm o} -\delta\tau_{\rm c}$ are of
the same  sign and of  similar magnitude as  that of $\delta\tau $ in
Figures 2 and  4.   This suggests that the simple amplitude-boosting correction
scheme presented here reduces, to some extent,  the systematic shift in the travel times
caused by the reduction of oscillation amplitudes in the sunspot.  Possible complications
for the  correction scheme include the spatial averaging that was used to create the scale-up function
$S_{\rm f}$, noise in the estimate of the amplitude suppression function $A(\mathbf x)$,  and
frequency dependence of the real solar suppression of oscillation amplitudes.
A more detailed  analysis of this correction scheme and
also some variants of it, including  a study of how the corrections in
travel times  affect the subsurface inferences  through inversions, is
left for a separate paper (Paper II) \citep{zhaoetal06}.

\section{Discussion}
\label{sec:discuss}

The  change  in  travel  times  in  response  to  changes  in  surface
oscillation amplitudes depends on  the spatial gradient of the amplitude modulation,
the amount of reduction  in the amplitudes, the travel distances, and the details of the 
phase-speed filter.
The  principal finding  is that  the largest  and  significant changes
occur only for waves with  short travel distances ($\Delta$ up to about
16 Mm). Values in the range of 5 - 40\%  of the travel-time anomalies that sunspots
cause could be a result of oscillation amplitude reduction (Figure 5b).
This  might indicate  that  the  subsurface
inferences from inversions  would, correspondingly, undergo significant
changes  for the  near-surface  layers. However,  the exact  amount of
changes  and how  the particular  $\Delta$ dependence  of $\delta\tau$
that we have shown here  would influence the inferences regarding deeper layers
can only be assessed by doing detailed analyses of inversions. We have
shown that a  simple correction, which involves boosting  up the p-mode
amplitudes,  is   able  to   reverse  the  interaction   of  amplitude
suppression and  phase-speed filtering  thereby removing substantially
the systematic  changes or  artifacts in travel  times. We  have shown
this for  the case  of a  small sunspot, where  there is  a measurable
amount of p-mode power within  the umbra.  In  large and very dark
sunspots, the  signal-to-noise ratio for the  p-mode signal in the umbrae
is too low to carry  out this correction successfully.

We  have  demonstrated  that  the  effects  of  oscillation  amplitude
variations on travel times are caused by the phase-speed filtering
procedure, which is however  crucial to achieving high signal-to-noise
as  well as  high spatial  resolution  in the  measurements of  travel
times.  Spatial amplitude  modulations (convolutions in Fourier space)
and  the phase-speed  filtering  are non-commuting  operations in  the
frequency-wavenumber  (Fourier)  domain.   The  travel  distance
$\Delta$  dependence  of  the   systematic  changes  in  travel  times
$\delta\tau$ are  seen to be of the  same form as the  actual travel-time
anomalies measured over sunspots (compare Figures 3b and 5a).
In spite of such a similarity  between the systematic errors and the
real changes  in $\delta\tau$ for  sunspots, it is important  that the
other  known  signatures that sunspots  leave  in  local
helioseismic  measurements  are  differentiated  from the  above. 
In particular,  sunspots show large asymmetries 
measured  in both the amplitudes of  cross-covariances and travel
times  \citep{duvalletal96},   as  well  as  in the  control-correlation
ingression   and  egression   measurements of helioseismic holography \citep{lindseyandbraun05},
between  the  out- and  in-going  wave  correlations.
These asymmetries possibly relate to
the irreversible changes in acoustic waves impinging on real sunspots and
hence their origin  is independent  of the travel-time shifts that  we have
shown here.  The contributions due to the effect that we have studied
here  are also  likely to  be present  in the  helioseismic holography
studies \citep{lindseyandbraun97,lindseyandbraun05}; because these studies do involve
selecting in Fourier space modes  of certain frequency-wavenumber
range, and hence a similar influence as that of a phase-speed filter is
possible.

\acknowledgments 
This  work utilizes data from  the Solar Oscillations
Investigation/  Michelson Doppler  Imager (SOI/MDI)  on the  Solar and
Heliospheric Observatory (SOHO). The  MDI project is supported by NASA
grant  NAG5-13261  to  Stanford   University.  SOHO  is  a  project  of
international cooperation  between ESA and  NASA. 
The work in part was supported by the UK Particle Physics and Astronomy
Research Council (PPARC) through grants PPA/G/S/2000/00502, 
PPA/V/S/2000/00512 and PP/X501812/1.
The work of  ACB was
supported by NASA contract NNH04CC05C. SPR thanks Dr. A.G. Kosovichev
for discussions and critical comments. We thank Dr. Douglas Braun
for useful comments.

\newpage

\clearpage

\clearpage

\end{document}